\documentclass[nofootinbib,twocolumn,showpacs,preprintnumbers,amsmath,amssymb,superscriptaddress,aps,pra,10pt]{revtex4-1}

\usepackage{amsmath}
\usepackage{amssymb}
\usepackage{amsfonts}
\usepackage{amsthm}
\usepackage{graphicx}

\makeatletter

\newcommand{\ket}[1]{|{#1}\rangle}
\newcommand{\kb}[1]{|{#1}\rangle\!\langle{#1}|}

\newcommand{\bra}[1]{\langle{#1}|}
\newcommand{\bkt}[2]{\langle{#1}|{#2}\rangle}

\newcommand{\Aw}{\langle \hat{A} \rangle _{w}}
\newcommand{\Atw}{\langle \hat{A}_{t} \rangle _{w}}
\newcommand{\nps}{{\rm nps}}
\newcommand{\ps}{{\rm ps}}
\newcommand{\erf}[1]{{\rm erf}\left[{#1}\right]}

\begin{document}
\title{Physical description of statistical hypothesis testing for a weak-value-amplification experiment using a birefringent crystal}
\author{Yuki Susa}
\email{susa@th.phys.titech.ac.jp}
\affiliation{Department of Physics, Tokyo Institute of Technology, Tokyo {\it 152-8551}, Japan}
\date{\today}
\begin{abstract}
We investigate the weak measurement experiment demonstrated by Ritchie {\it et al.} [N. W. M. Ritchie, J. G. Story, and R. G. Hulet, Phys. Rev. Lett. {\bf 66}, 1107 (1991)] from the viewpoint of the statistical hypothesis testing for the weak-value amplification proposed by Susa and Tanaka [Y. Susa and S. Tanaka, Phys. Rev. A {\bf 92}, 012112 (2015)].
We conclude that the weak-value amplification is a better method to determine whether the crystal used in the experiment is birefringent than the measurement without postselection, when the angles of two polarizers are almost orthogonal.
This result gives a physical description and intuition of the hypothesis testing and supports the experimental usefulness of the weak-value amplification. 
\end{abstract}

\maketitle

\section{Introduction}
In 1988, the weak measurement was proposed by Aharonov {\it et al.} as an indirect quantum measurement with the postselection of the measured system~\cite{Aharonov1988}.
Many theoretical and experimental studies have been done for the weak measurement in recent years~\cite{Dressel2014}.
Some researchers focused on the usefulness of the weak measurement as a technique for a high-precision measurement~\cite{Steinberg2010}.
From the weak measurement, we can extract a weak value which can be outside the range of the eigenvalues~\cite{Duck1989} or even complex.
The weak value is defined by 
\begin{align}
\Aw=\frac{\bra{f}\hat{A}\ket{i}}{\bkt{f}{i}},
\end{align}
where $\hat{A}$ is an observable and $\ket{i}$ and $\ket{f}$ are the initial and the final states of the measured system, respectively.
The weak measurement magnifies the output more than the one given by an ordinary projective measurement.
The weak value appears as a shift of the probe wave function induced by an interaction between the measured system and the measuring probe after postselecting the final state of the measured system~\cite{Jozsa2007}.

Actually, several experiments confirmed the weak-value amplification (WVA) effect~\cite{Ritchie1991, Hosten2008, Dixon2009}.
Some theoretical papers have shown that WVA is robust against systematic or technical error~\cite{Jordan2014, Lee2014}.
On the other hand, there is a statistical argument that WVA has a disadvantage in the parameter estimation for the interaction strength, because the postselection makes the number of detectable data small~\cite{Knee2013,Tanaka2013,Ferrie2014,Knee2014,Knee2014_2}. The countercriticism also arose that the data loss by postselection is not critical in practical cases~\cite{Vaidman2014}.

Reference~\cite{Susa2015} shows that WVA can be more significant for interaction detection than the measurement without postselection when the weak value is outside the eigenvalue range of the measured system observable.
The WVA is more likely than the ordinary measurement to correctly indicate when the interaction indeed exists.
This fact is derived by hypothesis testing~\cite{Ferguson1967}, which is one of the methods of statistical inference.
There it is supposed that the interaction is present if the measured value sufficiently deviates from the initial probe fluctuation.
The decision method gives a uniformly most powerful unbiased (UMPU) test, i.e., a statistically good testing.

In this work we demonstrate the testing for WVA~\cite{Susa2015} by comparing the power defined in the statistical hypothesis testing in two measurements, the weak measurement and the ordinary one without postselection, for a particular experimental setup.
We pick up the classic experiment using a birefringent crystal and two polarizers demonstrated by Ritchie {\it et al.}~\cite{Ritchie1991} for example.
This experiment was originally designed for the measurement of the weak value.
Here we look at the same experiment from a different angle.
We regard it as a testing problem to distinguish whether the crystal is birefringent or not.
Thus the statistical power in this experimental setup is given as the probability to determine exactly when the crystal is indeed birefringent.
This experiment gives intensity distributions as the results of the weak measurement including the case that the weak-coupling approximation~\cite{Aharonov1988} does not hold and the ordinary one with the birefringent crystal.
These experimental results clearly show the physical intuition of the testing and the advantage of WVA by comparing the powers as presented in Ref.~\cite{Susa2015}.
We conclude that the angle of the polarizers that give the weak value is the only factor in determining the case that the weak measurement is superior.
For other more recent WVA experiments, see the discussion in Sec. IV.

This paper is structured as follows.
In Sec. II we introduce the weak measurement of the experiment presented by Ritchie {\it et al.}~\cite{Ritchie1991}. 
In Sec. III we implement the hypothesis testing to decide whether the crystal used in the experiment is birefringent or not.
In Sec. IV we summarize this paper and discuss the issue of the number of detectable data in the actual case.
Some equations are provided in the Appendix A for convenience.

\section{Review of birefringence experiment for weak measurement}
Here we briefly review the optical setup of the experiment by Ritchie {\it et al.}~\cite{Ritchie1991}.
In this experiment the position $y$ of the laser beam is the measuring probe and the polarization is regarded as the measured system.
We use the Gaussian shape for the laser beam profile given by
\begin{align}
\psi(y)=\bkt{y}{\psi}=\left(\frac{1}{2\pi w_{0}^2}\right)^{\frac{1}{4}}e^{-\frac{y^2}{4w_{0}^2}}.
\end{align}
The waist of the beam used in the experiment~\cite{Ritchie1991} is $w_{0}=55\,\mu$m.
The tunings of the two polarizers play the roles of the pre- and postselection of the initial and final states in the measured system, which are
\begin{align}
\ket{i}=\cos \alpha \ket{H}+\sin \alpha \ket{V}, \\
\notag \\
\ket{f}=\cos \beta \ket{H}+\sin \beta \ket{V},
\end{align}
where $\ket{H}$ and $\ket{V}$ denote the horizontal and vertical polarization states, respectively, and
$\alpha$ and $\beta$ represent the angles of the first and second polarizers, respectively.

\begin{figure}[tp]
 \begin{center}
  \includegraphics[width=8.66cm]{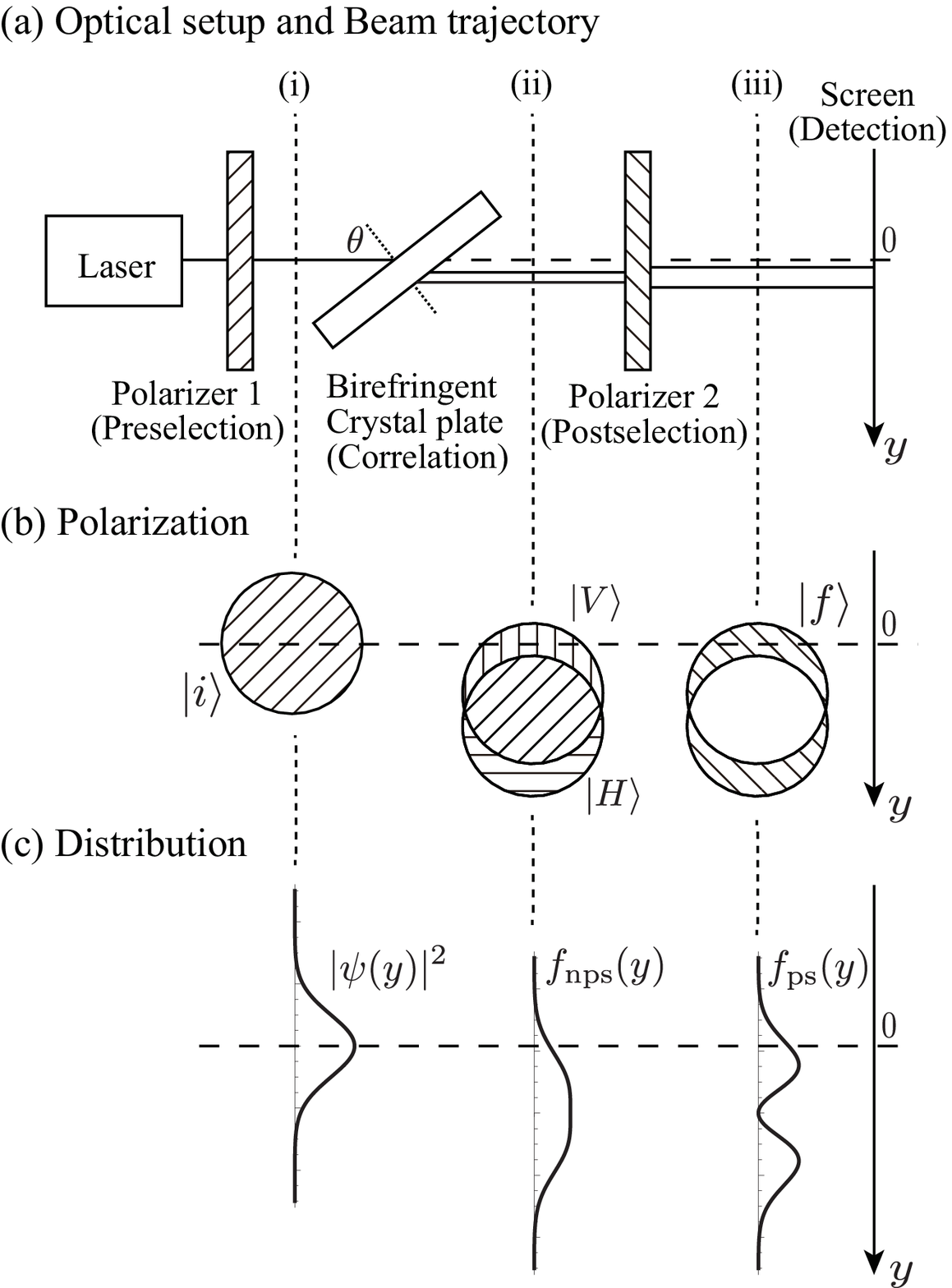}
 \end{center}
 \caption{(a) Sketch of the optical setup. The tilted birefringent crystal plates is set between the two polarizers, the angles of which are tuned almost orthogonal. On the screen, we observe the beam position $y$ and we decide whether or not the crystal is birefringent. Also shown are (b) the beam polarization and (c) the probe distribution, in each stage. For the illustration, the angles of the polarizers are $\alpha=\pi/4$ and $\beta=3\pi/4$.}
 \label{fig}
\end{figure}

The experimental setup is displayed in Fig.~\ref{fig}. 
Column (i) in Fig.~\ref{fig} shows the initial state of the polarization and the probe distribution. 
Passing through the first polarizer, the photons are injected into the birefringent crystal, which gives two different refraction factors depending on the polarization of the injected photons.
The injected beam is spatially separated into two beams with different polarizations, one of which is called an ordinary ray and the other is an extraordinary ray.
Then it gives the correlation between the position of the beam and the polarization. 
The refraction of the photons is described by the von Neumann-type Hamiltonian as
\begin{align}
\label{eq:hamiltonian}
\hat{H}=g\delta(t)\hat{A}\otimes \hat{p}_y,
\end{align}
where $g$ is the interaction strength, $\hat{A}$ is the observable given below, and $\hat{p}_y$ is the momentum operator conjugate to the position $y$ of the measuring probe.
Because the refraction factor depends on the polarization, $\hat{A}$ is given by
\begin{align}
\hat{A}=\lambda_{H}\kb{H}+\lambda_{V}\kb{V},
\end{align}
where $\lambda_{H,\,V}>0$ are the eigenvalues of $\hat{A}$.
We note that the crystal used in the experiment is a quartz plate, the refraction factors of which are $n_e=1.551\,65$ for the extraordinary ray and $n_o=1.542\,61$ for the ordinary ray when the wavelength of the injected laser is 633\,nm as quoted by Ref.~\cite{Ritchie1991}.
We also note that if the crystal is not birefringent, $\lambda_{H}=\lambda_{V}$.

The distribution function of the measuring probe after refraction is calculated as
\begin{align}
\label{eq:fnps}
&f_{\nps}(y|\lambda_{H}, \lambda_{V}) \notag \\
&~=|\bra{y}e^{-ig\hat{A}\otimes\hat{p}_y}\ket{\psi}\ket{i}|^2 \notag \\
&~=\frac{1}{\sqrt{2\pi w_{0}^{2}}}\Big(\cos^{2}\alpha\ e^{-\frac{(y-g\lambda_{H})^2}{2w_{0}^{2}}}+\sin^{2} \alpha\ e^{-\frac{(y-g\lambda_{V})^2}{2w_{0}^{2}}} \Big).
\end{align}
This output is obtained by the ordinary measurement without postselection.
As we can see, it is composed of the two Gaussian distributions.

According to Ref.~\cite{Ritchie1991}, the crystal thickness is $d=331\,{\mu}$m and the angle of incidence is $\theta\approx30^\circ$.
Using Snell's law, we can associate $g\lambda_{H}$ and $g\lambda_{V}$ with the position shifts by the refractions as
\begin{align}
g\lambda_{H}&=d\frac{\sin(\theta-\theta_{e})}{\cos \theta_{e}}\approx 67.92\,\mu {\rm m}, \\
g\lambda_{V}&=d\frac{\sin(\theta-\theta_{o})}{\cos \theta_{o}}\approx 67.28\,\mu {\rm m},
\end{align}
where $\sin \theta_e=\sin \theta / n_e$ and $\sin \theta_o=\sin \theta / n_o$.
Here we have regarded the horizontal and vertical polarized beams as extraordinary and ordinary rays, respectively.
We see that the difference of 0.64\,${\mu}$m between $g\lambda_{H}$ and $g\lambda_{V}$ in the birefringence is much smaller than the beam waist $w_{0}=55\,\mu$m.
Because the two beams almost overlap,
we would observe a single-peak distribution, the median of which is $y=g(\lambda_{H}+\lambda_{V})/2$, as the final probe distribution obtained by the photodetector.
Then it is difficult to distinguish whether or not the crystal is birefringent [Fig.~\ref{fig}(c)(ii)].
We note that the polarization of the overlapped region is the same as the initial polarization,
while the polarization of the nonoverlapped region is as shown in Fig.~\ref{fig}(b)(ii).

The probability distribution will be changed from $f_{\nps}(y)$ due to postselection by the second polarizer. 
The distribution function obtained by the weak measurement becomes
\begin{widetext}
\begin{align}
\label{eq:fps}
&f_{\ps}(y|\lambda_{H}, \lambda_{V}) \notag \\
&~=\frac{|\bra{y}\bra{f}e^{-ig\hat{A}\otimes\hat{p}_y}\ket{i}\ket{\psi}|^2}{|\bra{f}e^{-ig\hat{A}\otimes\hat{p}_y}\ket{i}\ket{\psi}|^2} \notag \\
&~=\frac{\cos^{2}\alpha \cos^{2} \beta\ e^{-\frac{(y-g\lambda_{H})^2}{2w_{0}^{2}}}+\sin^{2} \alpha \sin^{2}\beta\ e^{-\frac{(y-g\lambda_{V})^2}{2w_{0}^{2}}}+\frac{1}{2}\sin 2\alpha \sin 2\beta\ e^{-\frac{1}{2w_{0}^{2}}\left\{y- g\left(\frac{\lambda_{H}+\lambda_{V}}{2}\right)\right\}^{2}-\frac{g^{2}}{2w_{0}^{2}}\left(\frac{\lambda_H -\lambda_V }{2}\right)^2 } }{\sqrt{2\pi w_{0}^{2}}\Big[ \cos^{2}\alpha \cos^{2} \beta+\sin^{2}\alpha \sin^{2}\beta +\frac{1}{2}\sin 2\alpha \sin 2\beta\ e^{-\frac{g^{2}}{2w_{0}^{2}}\left(\frac{\lambda_H -\lambda_V }{2}\right)^2} \Big]}.
\end{align}
\end{widetext}
The third term of the numerator represents the interference.
We remark that the coefficient $\sin 2\alpha \sin 2\beta$ can be negative when the angles of the two polarizers are nearly orthogonal.
Its negativity tends to depress the final probe distribution in the central region and separate it into the two-peak distribution, because the polarization of the central part just after the crystal is orthogonal to the one of the second polarizer, which is used as the postselection [Fig.~\ref{fig}(b)(ii)].
The weak value becomes
\begin{align}
\Aw=\left(\frac{\lambda_{H}-\lambda_{V}}{2}\right)\frac{\cos(\alpha+\beta)}{\cos(\alpha-\beta)}+\frac{\lambda_{H}+\lambda_{V}}{2}.
\end{align}
When $\lambda_{H}\neq\lambda_{V}$, we can see that the weak value will be large for the almost orthogonal pair of $\alpha$ and $\beta$.
Then we could observe the amplified peak-to-peak distance by using WVA [Fig.~\ref{fig}(c)(iii)] to conclude that the crystal is birefringent.
However, the disadvantage of the postselection is the decrease of the entire intensity.

\section{Hypothesis testing for birefringence}
As an application we consider the statistical hypothesis testing proposed in Ref.~\cite{Susa2015} in the experiment that uses the birefringent crystal as explained in the previous section.
For the testing problem to determine whether the crystal is birefringent or not, we take the following hypotheses: the null hypothesis $H_0$, in which the crystal is not birefringent (i.e., $\lambda_{H}=\lambda_{V}$), and the alternative hypothesis $H_1$, in which the crystal is birefringent (i.e., $\lambda_{H}\neq\lambda_{V}$). 
We note that by the interaction, the refraction occurs for both hypotheses, which somewhat differs from the previous work~\cite{Susa2015}.

We compare the testing power given by the weak measurement and the one given by the measurement without postselection.
The testing power is defined as
\begin{align}
\label{eq:power}
b(\lambda_{H}, \lambda_{V}):=\int d(y) f(y|\lambda_{H}, \lambda_{V}) dy,
\end{align}
where $d(y)$ is a decision function which is a mathematical expression for a decision criterion.
The function $d(y)$ takes a binary value of 0 or 1.
The $0$ indicates that the null hypothesis is accepted and the $1$ represents that the alternative one is accepted.
The power (\ref{eq:power}) with $\lambda_{H}\neq\lambda_{V}$ indicates the probability to correctly judge that the alternative hypothesis is actually true.

In the previous research~\cite{Susa2015} that treats the testing problem to determine whether the interaction is present ($g\neq0$) or absent ($g=0$) between the measured system and the measuring probe,
the proposed decision function
\begin{align}
\label{eq:test}
d (y) &=\left\{
\begin{array}{l}
0,~~\text{if} ~~ |y|/w_{0}<c\\
1,~~\text{if} ~~ |y|/w_{0}\geq c
\end{array}
\right.
\end{align}
works well. 
Here $c$ is a critical point indicating the threshold beyond which the null hypothesis is rejected.
The physical interpretation of the decision function (\ref{eq:test}) is that the interaction would be present if the observed position $y$ is outside the initial laser beam waist $w_0$.

However, this decision function does not suit the present birefringence testing problem as it is because the refraction could make the beam position $y$ shifted outside the initial beam waist $w_{0}$, although the null hypothesis is actually true.
Then, for the proper testing, we have to adjust the final probe wave function by shifting $\lambda_{+}:=(\lambda_{H}+\lambda_{V})/2$.
When the null hypothesis is true, for example, $g\lambda_{+}$ is the medium of a single Gaussian distribution for the final probe state.
On the other hand, when the alternative hypothesis is true, $g\lambda_{+}$ coincides with the mean of the two peaks of the final probe distribution.
We can grasp the value of $g\lambda_{+}$ by the preparatory experiment without postselection to just monitor the refraction by the crystal.
Thus, the final probe distribution can be adjusted by a translation such as $y\rightarrow y+g\lambda_{+}$.

We remark that this adjustment can be described as the unitary operator $\exp[ig\lambda_{+} \hat{I}\otimes \hat{p}_y ]$, where $\hat{I}$ is the identity operator of the measured system.
Then, the total unitary operator combining the two unitary operators given by the Hamiltonian (\ref{eq:hamiltonian}) and by the adjustment becomes
\begin{align}
&\hat{U}_{t}=\exp[ig\lambda_{+} \hat{I}\otimes \hat{p}_y ]\exp[-ig\hat{A}\otimes \hat{p}_y ] \notag \\
&=\exp[-ig(\hat{A}-\lambda_{+} \hat{I})\otimes \hat{p}_y ]=\exp[-ig\hat{A}_{t}\otimes \hat{p}_y].
\end{align}
Here we have introduced the total observable $\hat{A}_{t}:=\lambda_{-}\left(\kb{H}- \kb{V}\right)$ and its eigenvalue $\lambda_{-}:=(\lambda_{H}-\lambda_{V})/2$ for convenience. 
Now we can rewrite our two hypotheses as $H_0:\lambda_{-}=0$ and $H_1: \lambda_{-}\neq0$ and the total weak value as
\begin{align}
\Atw=\frac{\bra{f}\hat{A}_{t}\ket{i}}{\bkt{f}{i}}=\lambda_{-}\frac{\cos(\alpha+\beta)}{\cos(\alpha-\beta)}.
\end{align}

To calculate the testing powers of each measurement, we introduce (in the Appendix A) the adjusted distributions: Eq. (\ref{eq:fnpsc}) for the ordinary measurement without postselection and Eq. (\ref{eq:fpsc}) for the weak measurement.
In the previous study (see Sec. III B in Ref.~\cite{Susa2015}) we showed that the decision function (\ref{eq:test}) becomes the UMPU test for the previous testing problem of $H_{0}: g=0$ and $H_{1}: g\neq0$, while the present testing problem is $H_{0}: \lambda_{-}=0$ and $H_{1}: \lambda_{-}\neq0$.
Noting that the distribution functions (\ref{eq:fnpsc}) and (\ref{eq:fpsc}) are functions of the product $g\lambda_{-}$, we conclude that the decision function (\ref{eq:test}) is also the UMPU test for the current problem.

\begin{table}[tp]
\caption{Condition $C(\alpha, \beta)$ and the weak value are displayed with the assigned values of $\alpha$ and $\beta$ used in the actual experiment~\cite{Ritchie1991}. In each case, $\alpha=\pi/4$.}
\begin{tabular}{cccc}
\hline\hline \vspace{-8pt}\\
\vspace{4pt}& $\displaystyle \beta$&  $\displaystyle C(\alpha,\beta)$ &$\displaystyle \left|\Atw/\lambda_{-}\right|^{2}$ \\ \hline
(a)&$\pi/4$& 1 & 0\\ 
(b)&$3\pi/4+2.2\times10^{-2}$& $-0.999$ & 2065\\ 
(c)&$3\pi/4$& $-1$& indeterminate\\
\hline\hline
\end{tabular}
\label{table}
\end{table}

The testing power of the measurement without postselection $b_{\nps}(\lambda_{-})$ is given by Eq. (\ref{eq:power_nps}) and the one of the weak measurement $b_{\ps}(\lambda_{-})$ is Eq. (\ref{eq:power_ps}).
From these powers we obtain the relation (\ref{eq:relation}), which gives the inequality
\begin{align}
\label{eq:ineq}
b_{\ps}(\lambda_{-})\geq b_{\nps}(\lambda_{-}),
\end{align}
which holds under the condition that the pair of $\alpha$ and $\beta$ satisfies
\begin{align}
\label{eq:cond}
C(\alpha, \beta):=\sin2\alpha \sin2\beta \leq0.
\end{align}
Here we have used the inequality (\ref{eq:ineq_erf}).
As stated in Sec. II, the amplification effect is induced if this condition (\ref{eq:cond}) is satisfied.
We also remark that condition (\ref{eq:cond}) is related to the requirement for a weak value derived in Ref.~\cite{Susa2015} as $|\Atw|^{2}\geq \left|\lambda_{-} \right|^{2} \Leftrightarrow C(\alpha, \beta) \leq0.$

\begin{figure}[tp]
 \begin{center}
  \includegraphics[width=8.66cm]{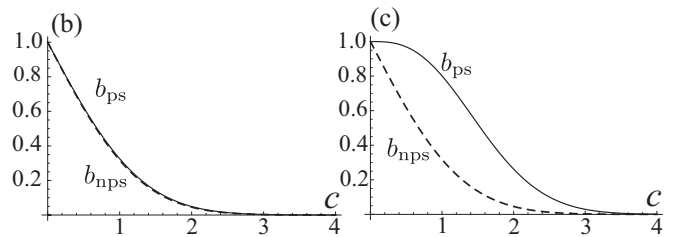}
 \end{center}
 \caption{Powers $b_{\rm ps}$ (solid line) and $b_{\rm nps}$ (dashed line) plotted with the critical points $c$ as the horizontal axis in the two cases: (b) $\beta=3\pi/4+2.2\times10^{-2}$ and (c) $\beta=3\pi/4$. The other parameters are fixed as $\alpha=\pi/4$, $w_{0}=55\,\mu$m, and $g\lambda_{-}=0.32\,\mu$m.}
 \label{fig2}
\end{figure}

The experimental results are summarized in Fig. 2 in Ref.~\cite{Ritchie1991}.
Table \ref{table} shows the value of $C(\alpha, \beta)$ and the weak value $|\Atw/\lambda_{-}|^{2}$ that the values of $\alpha$ and $\beta$ which were tuned in the actual experiment in the three cases. 
We focus on two cases: (a) the measurement {\it de facto} without postselection\footnote{The second polarizer slightly changes only the polarizations of the unoverlapped region, but not the one of the overlapped region.} and (c) the weak measurement with the postselected state orthogonal to the preselected state.
Case (a) exhibits the single Gaussian distribution, so we cannot distinguish whether the crystal is birefringent or not.
However, in case (c), which shows the two-peak distribution, we can clearly recognize that the crystal is birefringent.
We note that because the weak measurement case (b) meets the approximation $g|\Aw|\ll1$ , the final probe distribution is virtually a single Gaussian distribution shifted from the initial state as considered in Ref.~\cite{Aharonov1988}.
Case (c) gives significant amplification, which does not satisfy the approximation~\cite{Duck1989}. 
In both cases, the inequality (\ref{eq:ineq}) holds for the condition (\ref{eq:cond}).
Actually, case (c) gives a remarkable amplification effect as shown in Fig. \ref{fig2}, whereas the plots of $b_{\rm ps}$ and $b_{\rm nps}$ are almost overlapped in case (b).
Thus we have shown that the angles of the two polarizers are the only factor that determines the case when the weak measurement is superior to the ordinary measurement without postselection in terms of the testing power.

\section{Summary and discussion}
We have shown that the weak measurement, i.e., the measurement with the postselection, can be more powerful than the measurement without postselection in the hypothesis testing to determine whether or not the crystal is birefringent.
When the (total) weak value given by the angles of the two polarizers is larger than the eigenvalues of the (total) observable, WVA has the advantage for the present problem.
In particular, the pair of angles $\alpha$ and $\beta$, which does not satisfy the approximation $g|\Aw|\ll1$, gives the really powerful testing.
According to the authors of Ref.~\cite{Duck1989}, the amplification effect is rather striking when the approximation breaks down.
Our conclusion obtained thorough statistical analysis supports their view on WVA.
Here we have essentially treated the testing problem for the eigenvalue ($H_{0}: \lambda_{-}=0$ and $H_{1}: \lambda_{-}\neq0$), not the interaction strength ($H_{0}: g=0$ and $H_{1}: g\neq0$) that was treated in the previous work~\cite{Susa2015}.
In either case, the testing function (\ref{eq:test}) gives the UMPU test and works well.

It is often argued that postselection reduces the number of detectable data, which is a statistical disadvantage of WVA~\cite{Knee2013,Tanaka2013,Ferrie2014,Knee2014,Knee2014_2}.
On the other hand, the hypothesis testing generally works even if the number of detected data is small~\cite{Susa2015}.
We emphasize that the experiment by Ritchie {\it et al.}~\cite{Ritchie1991} has actually shown the birefringence of the crystal by the postselection,
although the detected intensity is much smaller ($\sim$$10^{-5}$) than that of the ordinary measurement case (Fig. 2 in Ref. ~\cite{Ritchie1991}).
We note that we observe no data with a completely orthogonal pair of $\alpha$ and $\beta$ when the null hypothesis is really true.
Then, practically, it is important to keep $\alpha$ and $\beta$ almost orthogonal but not quite, while the approximation is not still satisfied.

In the current task we have studied the classic experiment~\cite{Ritchie1991}, regarding it as testing the birefringence of the crystal.
The experiment is a helpful example to consider the hypothesis testing with WVA~\cite{Susa2015} because it is investigated outside the validity of approximation, especially the case of postselection completely orthogonal to preselection.
The hypothesis testing method can be applied to other WVA experiments, for instance, the detection of the spin Hall effect of light~\cite{Hosten2008} and sensing the tilted mirror in the interferometer~\cite{Dixon2009}.
To clearly show the effectiveness of WVA in the experiments quoted above,
we need the data for the region where the approximation ($g|\Aw|\ll1$) breaks down and the data of the ordinary measurements to compare.
With those data, it would be interesting to see the effectiveness of WVA by applying our hypothesis testing method.
We remark that to apply the proposed hypothesis testing method as it is, several assumptions (considered in Ref.~\cite{Susa2015}) are needed: that the probe state is given by the Gaussian profile, the measured system is described as the two-quantum-state system, and the experiment is based on the measurement of the position or the real part of the weak value.
If we want to test by measuring the momentum or the imaginary part of the weak value, we should establish the appropriate testing function for the experiment.

\section*{Acknowledgment}
The author thanks Professor A. Hosoya for valuable comments and reading the manuscript.

\appendix
\begin{widetext}
\section{}
Here we provide some equations to derive the inequality (\ref{eq:ineq}).
From the distributions (\ref{eq:fnps}) and (\ref{eq:fps}) we can obtain the adjusted distributions as
\begin{align}
\label{eq:fnpsc}
f_{\nps}^{{\rm adj}}(y|\lambda_{-}) &=f_{\nps}(y+g\lambda_{+}|\lambda_{H}, \lambda_{V}) =\frac{1}{\sqrt{2\pi w_{0}^{2}}}\Big(\cos^{2}\alpha\ e^{-\frac{(y-g\lambda_{-})^2}{2w_{0}^{2}}}+\sin^{2} \alpha\ e^{-\frac{(y+g\lambda_{-})^2}{2w_{0}^{2}}} \Big), \\
\label{eq:fpsc}
f_{\ps}^{{\rm adj}}(y|\lambda_{-})&=f_{\ps}(y+g\lambda_{+}|\lambda_{H}, \lambda_{V}) \notag \\
&=\frac{\cos^{2}\alpha \cos^{2} \beta\ e^{-\frac{(y-g\lambda_{-})^2}{2w_{0}^{2}}}+\sin^{2} \alpha \sin^{2}\beta\ e^{-\frac{(y+g\lambda_{-})^2}{2w_{0}^{2}}}+\frac{1}{2}\sin 2\alpha \sin 2\beta\ e^{-\frac{y^2 + g^{2}\lambda_{-}^2}{2w_{0}^{2}}} }{\sqrt{2\pi w_{0}^{2}}\Big(\cos^{2}\alpha \cos^{2} \beta+\sin^{2}\alpha \sin^{2}\beta +\frac{1}{2}\sin 2\alpha \sin 2\beta\ e^{-\frac{g^{2}\lambda_{-}^2}{2w_{0}^{2}}}\Big)}.
\end{align}
On the basis of the decision function (\ref{eq:test}) and the adjusted distributions (\ref{eq:fnpsc}) and (\ref{eq:fpsc}),
we can calculate the testing power (\ref{eq:power}) of the measurement without postselection as
\begin{align}
\label{eq:power_nps}
b_{\nps}(\lambda_{-})=1-\frac{1}{2}\left(\erf{\frac{cw_0 - g\lambda_{-}}{\sqrt{2w_0^2}}} +\erf{\frac{cw_0 + g\lambda_{-}}{\sqrt{2w_0^2}}}\right)
\end{align}
and the one of the weak measurement as
\begin{align}
\label{eq:power_ps}
b_{\ps}(\lambda_{-})=1-\frac{(\cos^2\alpha \cos^2 \beta + \sin^2 \alpha \sin^2 \beta )\left(\erf{\frac{cw_0 - g\lambda_{-}}{\sqrt{2w_0^2}}} +\erf{\frac{cw_0 + g\lambda_{-}}{\sqrt{2w_0^2}}}\right)+\sin 2\alpha \sin 2\beta\ e^{-\frac{g^2 \lambda_{-}^2}{2w_0^2}}\erf{\frac{c}{\sqrt{2}}}}{2(\cos^2 \alpha \cos^2 \beta + \sin^2 \alpha \sin^2 \beta) + \sin 2\alpha \sin 2\beta\ e^{-\frac{g^2 \lambda_{-}^2}{2w_0^2}}}.
\end{align}
From these we obtain the relation 
\begin{align}
\label{eq:relation}
\frac{1-b_{\ps}(\lambda_{-})}{1-b_{\nps}(\lambda_{-})}-1=\frac{\sin 2\alpha \sin 2\beta\ e^{-\frac{g^2 \lambda_{-}^2}{2w_0^2}}\bigg(\frac{2\erf{c/\sqrt{2}}}{\erf{(cw_0 - g\lambda_{-})/\sqrt{2w_0^2}} +\erf{(cw_0 + g\lambda_{-})/\sqrt{2w_0^2}}}-1\bigg)}{ 2(\cos^2 \alpha \cos^2 \beta + \sin^2 \alpha \sin^2 \beta) + \sin 2\alpha \sin 2\beta\ e^{-\frac{g^2 \lambda_{-}^2}{2w_0^2}} }.
\end{align}
We can derive the inequality (\ref{eq:ineq}) for Eq. (\ref{eq:cond}) by using the inequality
\begin{align}
\label{eq:ineq_erf}
\frac{2\erf{c/\sqrt{2}}}{\erf{(cw_0 - g\lambda_{-})/\sqrt{2w_0^2}} +\erf{(cw_0 + g\lambda_{-})/\sqrt{2w_0^2}}}-1>0,
\end{align}
which holds for $g\lambda_{-}\neq0$ and is shown in Appendix A of Ref.~\cite{Susa2015}.
\vspace{1mm}
\end{widetext}

\end{document}